# Outage Analysis and Optimization for Time Switching-based Two-Way Relaying with Energy Harvesting Relay Node

**Guanyao Du[1,2], Ke Xiong[*1,2], Yu Zhang[3] and Zhengding Qiu[1]**
[1] School of Computer and Information Technology, Beijing Jiaotong University, China
[2] National Mobile Communications Research Laboratory, Southeast University, China
[3] School of Computer and Communication Engineering, University of Science and Technology Beijing, China
[e-mail: 08112076{ kxiong, 06112049, zdqiu}@bjtu.edu.cn]
*Corresponding author: Ke Xiong

*Abstract*

Energy harvesting (EH) and network coding (NC) have emerged as two promising technologies for future wireless networks. In this paper, we combine them together in a single system and then present a time switching-based network coding relaying (TSNCR) protocol for the two-way relay system, where an energy constrained relay harvests energy from the transmitted radio frequency (RF) signals from two sources, and then helps the two-way relay information exchange between the two sources with the consumption of the harvested energy. To evaluate the system performance, we derive an explicit expression of the outage probability for the proposed TSNCR protocol. In order to explore the system performance limit, we formulate an optimization problem to minimize the system outage probability. Since the problem is non-convex and cannot be directly solved, we design a genetic algorithm (GA)-based optimization algorithm for it. Numerical results validate our theoretical analysis and show that in such an EH two-way relay system, if NC is applied, the system outage probability can be greatly decreased. Moreover, it is shown that the relay position greatly affects the system performance of TSNCR, where relatively worse outage performance is achieved when the relay is placed in the middle of the two sources. This is the first time to observe such a phenomena in EH two-way relay systems.

This work was supported by the Fundamental Research Funds for the Central Universities (no. 2014JBM024), by the National Natural Science Foundation of China under Grants (no. 61201203) and also by the Open Research Fund of National Mobile Communications Research Laboratory, Southeast University (no. 2014D03).

## 1. Introduction

**R**ecently, energy harvesting (EH) has emerged as a promising approach to realize green communications. Compared with conventional EH sources (e.g., solar, wind, thermoelectric effects or other physical phenomena) [1-3], one encouraging way is to harvest energy from the ambient radio-frequency (RF) signals, which is also referred to as simultaneous wireless information and energy transfer (SWIET).

The concept of SWIET was first proposed in [4], which is based on the fact that RF signals can carry both energy and information simultaneously. Because most devices used in wireless network are surrounded by RF signals (e.g., Wi-Fi signals or cellular signals), the technology of SWIET is particularly suitable for cooperative communication networks, where the transmissions of cooperation nodes can be powered by the energy harvested from the incoming signals rather than external energy supply. So far, the concept of SWIET has been extended to various wireless systems, such as relaying systems [5-7], multiple-input multiple-output (MIMO) systems [8], orthogonal frequency division multiplexing (OFDM) systems [9], and cooperative transmission systems [10-13], etc.

However, only a few work has been done for the two-way relay networks. As is known, two-way relay systems are able to improve the system spectral efficiency and system throughput greatly compared with one-way relay model, especially when network coding (NC) is applied to it [14-17]. More recently, some work began to apply EH to the two-way relay system to improve the performance of energy-constrained systems, see e.g. [10-12]. Specifically, in [10], the short-term sum-rate maximization problem was considered in the two-way relay model. In [11], a robust beamforming design was considered to maximize the sum-rate in an amplify-and-forward (AF) based two-way relay system. Nevertheless, these studies only considered convectional EH technologies and the SWIET was not investigated in their work. Although in [12], it considered the SWIET for the two-way relay system and it derived the outage probability and the ergodic capacity for the system, only AF relaying protocol and the power splitting receiver architecture proposed in [18] were involved.

In this paper, we also investigate the two-way relay network with SWIET. Compared with previous work [12-13], some differences are stressed here. Firstly, we considered time switching receiver architecture (another practically reliable receiver) rather than the power splitting receiver architecture, because time-switching based EH receiver is relatively simpler than the power-splitting based EH receiver. Secondly, as the decode-and-forward (DF) relaying can achieve much better system performance compared with AF relaying especially in high SNR regime [19], we consider the DF relaying in our work. Thirdly, we aim to explore the outage performance limit for the system. Besides, we formulate optimization problems to seek the minimal outage probability for the system. Outage probability is one of the most important performance measures for the two-way relay model, and there have been lots of works investigating the outage performance in two-way relay systems [20-22].

The main contributions of this paper are summarized as follows:

- Firstly, we investigate the NC and EH together in a single system for the two-way relay networks and we design a time switching-based network coding relaying (TSNCR) protocol for it.
- Secondly, to evaluate the system performance, we derive an explicit expression of the outage probability for the proposed TSNCR protocol. To show the system performance gain brougt by NC, we also design a protocol by considering traditional four phases decode-and-forward relaying without NC, namely, the time switching-based four phases relaying (TSFPR) protocol, and then derive an explicit expression of its outage probability.
- Thirdly, in order to explore the system performance limit, we formulate two optimization problems to minimize the system outage probability for each proposed protocol. Since the problems are non-convex and cannot be directly solved, we design a genetic algorithm (GA)-based optimization algorithm for them.
- Finally, numerical results are provided to validate our theoretical analysis and show that in such an EH two-way relay system, if NC is applied, the system outage probability can be greatly decreased. Moreover, it is shown that the relay position greatly affects the system performance of TSNCR, where relatively worse outage performance is achieved when the relay is placed in the middle of the two sources. This is the first time to observe such a phenomena in EH two-way relay systems.

The rest of the paper is organized as follows. In Section 2, we present the system model and assumptions. In Section 3, we describe the proposed TSNCR and TSFPR protocols for the two-way transmission and analyze their corresponding system outage performance, respectively. In Section 4, the optimization problems are formulated and a GA-based algorithm is proposed to find the optimal system parameters to achieve the minimum system outage probability. In Section 5, we provide numerical results. Finally, the conclusion is followed in Section 6.

## 2. System Model

### 2.1 Assumptions and Notations

Consider a two-way relay wireless network composed of two sources, $U_1$ and $U_2$, and an energy harvesting relay $R$, as shown in **Fig. 1**. The two sources want to exchange their information with each other via $R$ through orthogonal channels. $h_i$ and $f_i$ are used to denote the channel gain of the links from $U_i$ to $R$ and from $R$ to $U_i$, respectively, where $i=1,2$. Quasi-block fading channel model is adopted here, following Rayleigh fading model. The transmit power and the distance from $U_i$ to $R$ are denoted by $P_i$ and $d_i$, respectively.

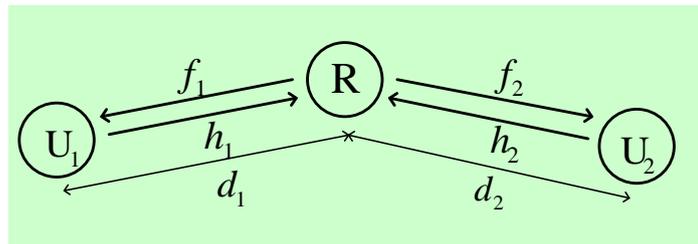

**Fig. 1** System model

In each round of information exchange, the sources transfer both information and energy to $R$ through its own transmitted RF signals simultaneously. The delivered energy can be obtained by $R$ from the recipients to recharge its battery, and then uses all the harvested energy to help the information exchange between the two sources[1]. Note that, the data buffer and the battery at the relay are assumed to be sufficiently large, so no data and energy overflow at R are required to be considered for the system. Moreover, half-duplex mode is assumed and all nodes in the system are equipped with a single antenna.

## 3. Relaying Protocols and Outage Performance Analysis

Based on these assumptions, in this section we first introduce the proposed network coding-based relaying protocol, and analyze the corresponding outage probability for it. Then for the purpose of comparing, we also present the four phases relaying protocol in which the relay operates in DF cooperative scheme.

### 3.1 Time Switching-Based Network Coding Relaying (TSNCR) Protocol

### 3.1.1 Protocol Description

As shown in **Fig. 2(a)**, for a time period $T$, $\rho T$ part is assigned for R to harvest energy from the two sources, where it is equally divided into two durations and the $i$-th $\rho T/2$ duration is assigned to R to harvest energy from the received signal from $U_i$. The remaining time $(1-\rho)T$ is used for the two-way information transmission, where it is equally divided into three durations and each duration is with $(1-\rho)T/3$ duration. During the first two $(1-\rho)T/3$ durations, $U_i$ transmits its information to R with power $P_i$ to R, and in the third $(1-\rho)T/3$ duration, R decodes the information received from $U_1$ and $U_2$, mixes the two flows of information with network coding (e.g., XOR coding operation) and then uses all the harvested energy to broadcast the network-coded information to the two sources. Once $U_1$ and $U_2$ received the broadcasted information from R, they can decode the desired information from the mixed information based on their prior knowledge.

**Figure. 2(b)** describes the relay receiver in the TS-based protocols. Each received RF signal $y_r$ from $U_i$ ($i=1,2$) at the relay is first sent to the energy harvesting receiver (for $\rho T/2$ time) and then to the information receiver (for $(1-\rho)T/3$ time for TSNCR). Note that $y_r$ is corrupted by two noises, where $n_{r,a}$ is introduced by the receiving antenna which is modeled as a narrowband Gaussian noise, and $n_{r,c}$ is the sampled additive noise due to the RF band to baseband signal conversion. The details of the EH receiver and the information receiver can be found in [18].

In the following subsection, we will analyze the system outage performance for the TSNCR protocol[2].

---

[1] which means that the harvested energy from $U_1$ and $U_2$ is exhausted to relay the information for the two users in the following phases within each round of two-way transmission.

[2] The signal processing energy cost is not considered in this paper, thus relaying transmission is the only energy operating cost for the system. Such an assumption is also widly adpoted in similar articles [5,12-13,23-24].

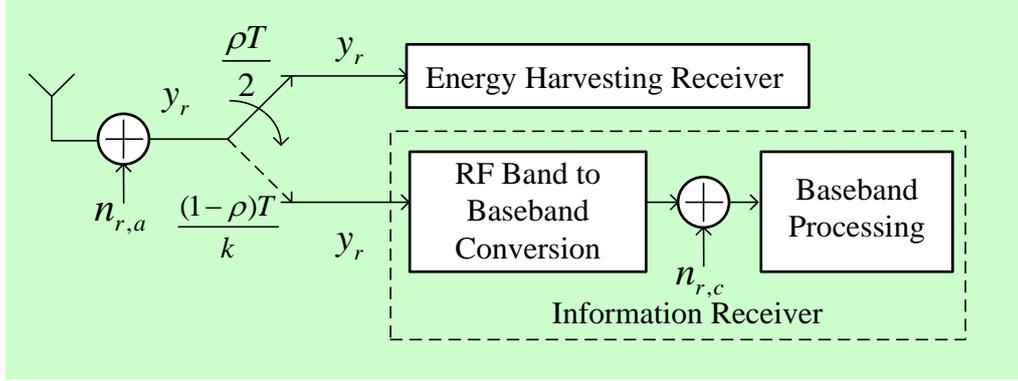

(a)

(b)

**Fig. 2** (a) TSNCR protocol (b) architecture of the relay receiver, where $k$ denotes the number of information transmissions in the proposed protocols, $k=3$ for TSNCR and $k=4$ for TSFPR.

### 3.1.2 Outage Probability Analysis for the TSNCR Protocol

As illustrated in **Fig. 2**, phase 1 and phase 2 are used for $U_1$ and $U_2$ to transfer energy and information to R respectively, and phase 3 is used for R to forward the received information to $U_1$ and $U_2$ with the harvested energy. During the $i$-th ($i=1,2$) phase, $U_i$ firsly transfers energy to R. Then the energy that R harvests from $U_i$ can be given by

$$E_{H,i} = \frac{\eta P_i |h_i|^2}{d_i^m}(\rho T/2), \qquad (1)$$

where $0 < \eta \leq 1$ is used to describe the energy conversion efficiency. Then $U_i$ transmits its information to R. So, at the end of the $i$-th phase, after some processing, the baseband signal at R is given by

$$y_{r,i} = \frac{1}{\sqrt{d_i^m}} \sqrt{P_i} h_i s_i + n_{r,a_i} + n_{r,c_i}, \qquad (2)$$

where $m$ denotes the path loss exponent, and $s_i$ is with unit average power. $n_{r,a_i}$ and $n_{r,c_i}$ denote the baseband additive white Gaussian noise (AWGN) with variance $\sigma_{r,a_i}^2$ introduced by the relay's receiving antenna and the sampled AWGN with variance $\sigma_{r,c_i}^2$ introduced by the conversion from RF band signal to baseband signal from $U_i$, respectively. The corresponding

data rate from $U_i$ at R is

$$R_{r,i} = \frac{1-\rho}{3} \log(1 + \frac{P_i |h_i|^2}{d_i^m \sigma_{r,i}^2}), \tag{3}$$

where $n_{r,i} = n_{r,a_i} + n_{r,c_i}$ with variance $\sigma_{r,i}^2 = \sigma_{r,a_i}^2 + \sigma_{r,c_i}^2$, which is actually the overall noise received at R in phase $i$.

After the first two phases, R has harvested total $E_{H,1} + E_{H,2}$ energy from the two sources. So, the transmit power at R in the third phase is

$$P_r = \frac{E_{H,1} + E_{H,2}}{(1-\rho)T/3} = \sum_{i=1}^{2} \frac{3\eta\rho P_i |h_i|^2}{2d_i^m(1-\rho)}. \tag{4}$$

Then, R decodes and re-encodes the received signals with NC and uses all the harvested energy to broadcast the coded information to $U_1$ and $U_2$ simultaneously. As $U_1$ and $U_2$ both know their own signal transmitted to R, they can decode the desired signal from the mixed information based on their prior knowledge. As a result, at the end of the relay broadcast phase, the received signal at $U_i$ ($i=1,2$) is given by

$$y_{u,i} = \frac{1}{\sqrt{d_i^m}} \sqrt{P_r} f_i s_j + n_{u,a_i} + n_{u,c_i}, \tag{5}$$

where $j=1,2$ and $j \neq i$. $n_{u,a_i}$ and $n_{u,c_i}$ are the baseband AWGN with variance $\sigma_{u,a_i}^2$ and the sampled AWGN with variance $\sigma_{u,c_i}^2$ at $U_i$, respectively. Therefore, the corresponding achievable date rate at $U_i$ can be given by

$$R_{u,i} = \frac{1-\rho}{3} \log(1 + \frac{P_r |f_i|^2}{d_i^m \sigma_{u,i}^2}), \tag{6}$$

where $\sigma_{u,i}^2 = \sigma_{u,a_i}^2 + \sigma_{u,c_i}^2$.

Since the broadcast data rate with digital network coding is bounded by the worse channel of $f_1$ and $f_2$, so the achievable transmission rate in the third phase is

$$R_r = \min(R_{u,1}, R_{u,2}) = \frac{1-\rho}{3} \log(1 + \text{SNR}_{bc}), \tag{7}$$

where $\text{SNR}_{bc} = \min\left(\frac{P_r |f_1|^2}{d_1^m \sigma_{u,1}^2}, \frac{P_r |f_2|^2}{d_2^m \sigma_{u,2}^2}\right)$.

As is known, the network coded two-way relay network is composed of three transmissions, i.e., the information delivering over $U_1 \to R$ link, the information delivering over $U_2 \to R$ link and the information broadcasted over $R \to U_i$. That is, any link's failure can lead to the occurrence of communication outage. According to [25], we define the system outage probability of TSNCR as

$$P_{out}^{(\text{TSNCR})} = 1 - \Pr[R_{r,1} \geq R_0] \cdot \Pr[R_{r,2} \geq R_0] \cdot \Pr[R_r \geq R_0]. \tag{8}$$

**Theorem 1**: *Given a target transmission rate $R_0$, the outage probability for the two-way relay network with EH relay in the TSNCR Protocol is given by (9), where $K_1(\cdot)$ and $K_2(\cdot)$ denote the first-order and second-order modified Bessel function of the second kind respectively* [26]. $a = 3\rho\eta P_1 d_1^{-m}(1-\rho)^{-1}/2$, $b = 3\rho\eta P_2 d_2^{-m}(1-\rho)^{-1}/2$, $a_0 = u_0 d_1^m \sigma_{r,1}^2 P_1^{-1}$, $b_0 = u_0 d_2^m \sigma_{r,2}^2 P_2^{-1}$,

$u_0 = 2^{3R_0/(1-\rho)} - 1$, $\quad e_0 = d_1^m \sigma_{u,1}^2 \lambda_{f_1} + d_2^m \sigma_{u,2}^2 \lambda_{f_2}$, $\quad \lambda_{f_1}$ and $\lambda_{f_2}$ are the mean value of the exponential random variables $|f_1|^2$ and $|f_2|^2$, respectively.

*Proof*: See the Appendix A. ∎

$$P_{out}^{(\text{TSNCR})} = \begin{cases} 1 - \dfrac{2\lambda_{h_1}\lambda_{h_2}}{a\lambda_{h_2} - b\lambda_{h_1}} e^{-\lambda_{h_1} a_0 - \lambda_{h_2} b_0} \left( \sqrt{\dfrac{ae_0 u_0}{\lambda_{h_1}}} K_1\left(2\sqrt{\dfrac{e_0 u_0 \lambda_{h_1}}{a}}\right) - \sqrt{\dfrac{be_0 u_0}{\lambda_{h_2}}} K_1\left(2\sqrt{\dfrac{e_0 u_0 \lambda_{h_2}}{b}}\right) \right) & a\lambda_{h_2} \neq b\lambda_{h_1} \\ 1 - \dfrac{2\lambda_{h_1} e_0 u_0}{a} e^{-\lambda_{h_1} a_0 - \lambda_{h_2} b_0} K_2\left(2\sqrt{\dfrac{e_0 u_0 \lambda_{h_2}}{b}}\right), & a\lambda_{h_2} = b\lambda_{h_1} \end{cases} \quad (9)$$

*Remark 1: With the system outage probability obtained in (9), the system outage capacity of the proposed TSNCR protocol for such a two-way relay network can be given as follows [5,22]:*

$$C_{out}^{(\text{TSNCR})} = (1 - P_{out}^{(\text{TSNCR})}) \frac{(1-\rho)T/3 \times 2}{T} \cdot R_0 = \frac{2}{3}(1-\rho)(1 - P_{out}^{(\text{TSNCR})})R_0, \quad (10)$$

*where the coefficient $2(1-\rho)/3$ results from the transmission duty cycle loss in TSNCR protocol for the two-way relay system.*

### 3.2 Time Switching-Based Four-Phase Relaying (TSFPR) Protocol

To compare the performance gain achieved by the combination of EH and NC technology, we also present the four phases protocol without NC relaying scheme, and analyze its outage probability.

#### 3.2.1 Protocol Description

Similar to TSNCR, we also use $\rho T$ part of the total time for R to harvest energy from the two sources, where each source is assigned $\rho T/2$ to transfer energy to R. In TSFPR, as the two-way relay information transmission is accomplished through four stages, the remaining $(1-\rho)T$ time that used for the information transmission is thus divided into four equal parts, where each is with $(1-\rho)T/4$. Since no network coding is used, R has to forward the received information to the two sources separately. Let $P_r$ is the available power at R to help forwarding the information. A part of the harvested energy $\theta P_r$ is used to forward the information to $U_1$, and the rest part of $(1-\theta)P_r$ is used to forward the information to $U_2$, where $0 \leq \theta \leq 1$ denotes the transmit power re-distribution factor at R. The architecture of the relay receiver in the TSFPR protocol is the same as the one in the TSNCR protocol which is depicted in **Fig. 2(b)**.

#### 3.2.2 Outage Probability Analysis for the TSFPR Protocol

Similar to the TSNCR protocol, at the end of the $i$-th ($i = 1, 2$) phase, the sampled baseband signal at R is given by

$$R_{r,i} = \frac{1-\rho}{4}\log(1+\frac{P_i|h_i|^2}{d_i^m \sigma_{r,i}^2}), \tag{11}$$

where $\sigma_{r,i}^2$ is defined below (3).

The energy that R harvests from $U_i$ is given by (1). So, the total transmit power harvested at R for the subsequent information relaying is given by

$$P_r = \frac{E_{H,1}+E_{H,2}}{(1-\rho)T/4} = \sum_{i=1}^{2}\frac{2\eta\rho P_i|h_i|^2}{d_i^m(1-\rho)}. \tag{12}$$

Then, R decodes and re-encodes the received signals and redistributes the total transmit power in $\theta:1-\theta$ portion, such that $\theta P_r$ is used for the information transmission to $U_1$ in the third phase, and the remaining part $(1-\theta)P_r$ is for the information relaying to $U_2$ in the fourth phase. Therefore, the corresponding date rate at $U_j$ ($j=1,2$) can be given by

$$R_{u,j} = \frac{1-\rho}{4}\log(1+\frac{\theta_j P_r |f_j|^2}{d_j^m \sigma_{u,j}^2}), \tag{13}$$

where $\theta_j = \theta$ for $j=1$ and $\theta_j = 1-\theta$ for $j=2$.

The outage probability $P_{out}^{(\text{TSFPR})}$ for the TSFPR protocol can be calculated as

$$P_{out}^{(\text{TSFPR})} = 1 - \Pr[R_{r,1}\geq R_0, R_{u,2}\geq R_0]\cdot \Pr[R_{r,2}\geq R_0, R_{u,1}\geq R_0]. \tag{14}$$

**Theorem 2**: *Given a target transmission rate $R_0$, the outage probability for the two-way relay network with EH relay in the TSFPR protocol can be analytically calculated using (15), where, $a = 2(1-\theta)\rho\eta P_1(1-\rho)^{-1}d_1^{-m}$, $b = 2(1-\theta)\rho\eta P_2(1-\rho)^{-1}d_2^{-m}$, $c = 2\theta\rho\eta P_1(1-\rho)^{-1}d_1^{-m}$, $d = 2\theta\rho\eta P_2(1-\rho)^{-1}d_2^{-m}$, $a_0 = (2^{4R_0/(1-\rho)}-1)d_1^m\sigma_{r,1}^2 P_1^{-1}$, $b_0 = (2^{4R_0/(1-\rho)}-1)d_2^m\sigma_{r,2}^2 P_2^{-1}$, $u_0 = 2^{4R_0/(1-\rho)}-1$, $c_0 = u_0 d_2^m \sigma_{u,2}^2$, $d_0 = u_0 d_1^m \sigma_{u,1}^2$, $e_0 = e^{-\lambda_{h_1}a_0 - \lambda_{h_2}b_0}$, $\lambda_{h_1}$, $\lambda_{h_2}$ $\lambda_{f_1}$, and $\lambda_{f_2}$, are the mean value of the exponential random variables $|h_1|^2$, $|h_2|^2$ $|f_1|^2$ and $|f_2|^2$, respectively.*

*Proof*: See the Appendix B. ∎

$$P_{out}^{(\text{TSFPR})} = \begin{cases} 1 - \frac{4\lambda_{h_1}^2\lambda_{h_2}^2 e_0}{(a\lambda_{h_2}-b\lambda_{h_1})(c\lambda_{h_2}-d\lambda_{h_1})}\left(\sqrt{\frac{c_0 a\lambda_{f_2}}{\lambda_{h_1}}}K_1\left(2\sqrt{\frac{c_0\lambda_{f_2}\lambda_{h_1}}{a}}\right) - \sqrt{\frac{c_0 b\lambda_{f_2}}{\lambda_{h_2}}}K_1\left(2\sqrt{\frac{c_0\lambda_{f_2}\lambda_{h_2}}{b}}\right)\right) \\ \quad \times \left(\sqrt{\frac{d_0 c\lambda_{f_1}}{\lambda_{h_1}}}K_1\left(2\sqrt{\frac{d_0\lambda_{h_1}\lambda_{f_1}}{c}}\right) - \sqrt{\frac{d_0 d\lambda_{f_1}}{\lambda_{h_2}}}K_1\left(2\sqrt{\frac{d_0\lambda_{h_2}\lambda_{f_1}}{d}}\right)\right), \quad a\lambda_{h_2}\neq b\lambda_{h_1},\ c\lambda_{h_2}\neq d\lambda_{h_1} \\ 1 - \frac{4\lambda_{h_1}^2\lambda_{h_2}\lambda_{f_1}d_0 e_0}{(a\lambda_{h_2}-b\lambda_{h_1})c}\left(\sqrt{\frac{c_0 a\lambda_{f_2}}{\lambda_{h_1}}}K_1\left(2\sqrt{\frac{c_0\lambda_{f_2}\lambda_{h_1}}{a}}\right) - \sqrt{\frac{c_0 b\lambda_{f_2}}{\lambda_{h_2}}}K_1\left(2\sqrt{\frac{c_0\lambda_{f_2}\lambda_{h_2}}{b}}\right)\right)K_2\left(2\sqrt{\frac{c_0\lambda_{h_2}\lambda_{f_1}}{d}}\right), a\lambda_{h_2}\neq b\lambda_{h_1},\ c\lambda_{h_2}=d\lambda_{h_1} \\ 1 - \frac{4\lambda_{h_1}^2\lambda_{h_2}\lambda_{f_2}c_0 e_0}{(c\lambda_{h_2}-d\lambda_{h_1})a}K_2\left(2\sqrt{\frac{c_0\lambda_{h_2}\lambda_{f_2}}{b}}\right)\left(\sqrt{\frac{d_0 c\lambda_{f_1}}{\lambda_{h_1}}}K_1\left(2\sqrt{\frac{d_0\lambda_{h_1}\lambda_{f_1}}{c}}\right) - \sqrt{\frac{d_0 d\lambda_{f_1}}{\lambda_{h_2}}}K_1\left(2\sqrt{\frac{d_0\lambda_{f_1}\lambda_{h_2}}{d}}\right)\right), a\lambda_{h_2}=b\lambda_{h_1},\ c\lambda_{h_2}\neq d\lambda_{h_1} \\ 1 - \frac{4\lambda_{h_1}^2\lambda_{f_2}\lambda_{f_1}c_0 d_0 e_0}{ac}K_2\left(2\sqrt{\frac{c_0\lambda_{h_2}\lambda_{f_2}}{b}}\right)K_2\left(2\sqrt{\frac{c_0\lambda_{h_2}\lambda_{f_1}}{d}}\right), \quad a\lambda_{h_2}=b\lambda_{h_1},\ c\lambda_{h_2}=d\lambda_{h_1} \end{cases} \tag{15}$$

**Remark 2**: *With the system outage probability obtained in (15), the system outage capacity of the proposed TSFPR protocol for such a two-way relay network can be given as follows [5,22]*:

$$C_{out}^{(\text{TSFPR})} = (1-P_{out}^{(\text{TSFPR})})\frac{(1-\rho)T/4\times 2}{T}\cdot R_0 = \frac{1}{2}(1-\rho)(1-P_{out}^{(\text{TSFPR})})R_0, \tag{16}$$

*where the coefficient $(1-\rho)/2$ results from the transmission duty cycle loss in TSFPR*

*protocol for the two-way relay system.*

## 4. Outage Performance Optimization

To explore the system performance limit, in this section, we shall formulate two optimization problems for the two protocols. Since both problems are non-convex, we also design a GA-based algorithm to solve them.

### 4.1 Optimization Problems

#### 4.1.1 Optimization Problem for TSNCR

For the TSNCR protocol, the system optimization problem can be formulated as follows

$$\min_{\rho} \; P_{out}^{(\text{TSNCR})}$$
$$\text{s.t.} \quad 0 \leq \rho \leq 1 \tag{17}$$

#### 4.1.2 Optimization Problem for the TSFPR

For the TSFPR protocol, the system optimization problem can be formulated as follows

$$\min_{\rho,\theta} \; P_{out}^{(\text{TSFPR})}$$
$$\text{s.t.} \quad 0 \leq \rho \leq 1, \; 0 \leq \theta \leq 1 \tag{18}$$

### 4.2 GA-based Algorithm

Because of the Bessel functions involved in the analytical expressions of $P_{out}$ shown in (9) and (15), it is difficult to derive an analytic solution of $\rho$ for TSNCR (and $\theta$ for TSFPR). Moreover, exhaustive search to get the optimal system configuration parameters is too inefficient to be executed in practice, especially for the TSFPR protocol in which multiple parameters need to be optimized. To solve this problem, in the following subsection, we shall design a GA-based optimization algorithm to obtain the optimal solution of system configuration parameters to achieve the optimal outage performance.

GA starts with the generation of a random population, which is a group of chromosomes. Each chromosome has a fitness which is evaluated against the objective function. According to the survivor selection criterion based on survival of the fittest, chromosomes with better fitness will survive for evolution while chromosomes with less fitness will be discarded. The evolution includes three operations: mate selection, crossover and mutation. Mate selection selects chromosome mates with better fitness from the survivors to create new offspring. Crossover is then executed over the selected chromosome mates to reproduce new offspring. Crossover is a process of gene recombination which can transfer partial genes from parents to offspring. Mutation is implemented to alter partial genes of offspring, which can avoid converging into local optimal solution fast. That is to say, new genes are generated after mutation, which leads to searching solutions in distinct area of solution space. The evolution process repeats until the termination conditions are satisfied [27][28].

Taking the TSFPR protocol for example, we shall show how the GA-based optimization algorithm works for it. By regarding $\rho$ and $\theta$ as genes respectively, the combination of $\rho$ and $\theta$ composes a chromosome. Objective function in (18) is used to calculate each chromosome's fitness. $Q_{\min}(t)$ denotes the optimal solution of the $t$-th generation, and $\delta$ is the predefined precision of GA. Main steps of the GA-based optimization algorithm are given as follows:

| | **GA-based Optimization Algorithm** |
|---|---|
| **Step 1** | *Population initialization*: $K_{\text{ini}}$ chromosomes are randomly generated as the initial population of the $t$-th ($t=0$) generation. |
| **Step 2** | *Fitness evaluation*: each chromosome's fitness is evaluated by objective function in (16). If $|Q_{\min}(t) - Q_{\min}(t-1)| < \delta$, algorithm ends; Otherwise, go to **Step 3**; |
| **Step 3** | *Survivor selection*: $K_{\text{sel}} = \varepsilon \cdot K_{\text{ini}}$ chromosomes with best fitness survive while the others with less fitness are eliminated to make room for new offspring, where $\varepsilon \in (0,1)$ is the selection rate to survival; |
| **Step 4** | *Mate selection*: Two mates are selected from the $K_{\text{sel}}$ survival chromosomes to produce new offspring by adopting Roulette wheel selection; |
| **Step 5** | *Crossover*: Single-point crossover is applied to produce two offspring from two selected mates. Sufficient offspring are produced until the number of survivors and offspring is equal to $K_{\text{ini}}$. |
| **Step 6** | *Mutation*: A probability $\mu$ is used to decide whether a gene is mutated or not. Gene mutation leads to the production of a new group of $\alpha$ and $\theta$. After mutation, $t = t+1$, and return to **Step 2**. |

When the GA-based optimization algorithm is applied to the TSNCR protocol, the chromosome is composed of only one gene $\rho$, and the objective function in (17) is used to calculate each chromosome's fitness.

The computational complexity of the proposed GA-based algorithm depends on the number of iterations $N_{\text{ite}}$ in GA ($N_{\text{ite}} \propto \delta^{-1}$), the number of chromosomes $K_{\text{ini}}$ in each iteration, and the complexity in evaluating the fitness value in (14), which has the computational complexity of $O(1)$. Therefore, the total complexity of our proposed method is $O(N_{\text{ite}} \cdot K_{\text{ini}})$. To the best of our knowledge, the convergence of GA for the case of finite iteration number is still an open problem [29]. So in this paper, instead of giving theoretical analysis, we shall investigate the convergence of GA-based optimization algorithm by simulations in Section 5.

## 5. Numerical Results

In this section, numerical results are provided to verify the theoretical analysis on the system outage probability for the two protocols and the effectiveness of the proposed GA-based algorithm. The effects of various system parameters on the system outage performance are also discussed, including $\rho$, $\theta$ (for the TSFPR protocol), source's transmit power and the relay location. we set $R_0 = 1$ bit/sec/Hz, $\eta = 1$, $P_1 = P_2 = 1$ Watt, $B = 1$ Hz, and $m = 2.7$ [5,23]. All the mean values of the exponential random variables $|f_1|^2$, $|h_1|^2$, $|f_2|^2$ and $|h_2|^2$ are set to be 1. For simplicity, we assume that $\sigma_a^2 = \sigma_{r,a_1}^2 = \sigma_{r,a_2}^2 = \sigma_{u,a_1}^2 = \sigma_{u,a_2}^2$, and $\sigma_c^2 = \sigma_{r,c_1}^2 = \sigma_{r,c_2}^2 = \sigma_{u,c_1}^2 = \sigma_{u,c_2}^2$. And in GA, $K_{\text{ini}} = 100$, $\varepsilon = 0.5$, $\mu = 0.05$ and $\delta = 10^{-5}$.

### 5.1 Verification of the Analytical Outage Probability

In **Fig. 3**, simulation results obtained through the Monte Carlo simulation using (8) and (14) are compared with our analytical expressions for the system outage probability developed through (9) and (15). It can be seen that, the analytical and the simulation results match well for the two protocols, which verifies the analytical expressions for $P_{out}$ presented in Theorem 1 and Theorem 2.

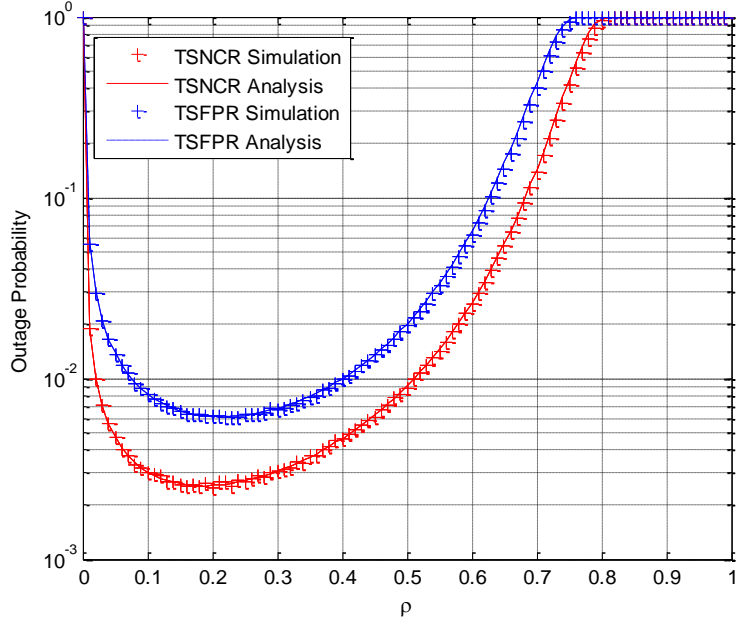

**Fig. 3** Outage probability: numerical vs simulation with $\theta = 0.5$.

## 5.2 Effect of $\rho$ and $\theta$ on System Outage Probability

**Figure 4** plots the optimal system outage probability versus $\rho$ and $\theta$ (only for TSFPR). For the TSNCR, the optimal outage probability first decreases and then increases with the increment of $\rho$, and it achieves the minimum at $\rho = 0.19$. For the TSFPR, there is another parameter $\theta$ which can also affect $P_{out}$. It can be observed that, no matter what value of $\theta$ is, the optimal outage probability firstly decreases with the increment of $\rho$ from 0 to the optimal value ($\rho = 0.22$), and then increases with the growth of $\rho$. This is due to the fact that, the relay obtains less transmit power ($P_r$) from energy harvesting for smaller $\rho$ than the optimal value, which incurs more outages in the relay cooperation phase. On the other hand, when $\rho$ gets higher, the relay may obtain more transmit power than the optimal value. Thus, less power is left for the users to transmit their own information to the relay. Consequently, poor signal strength is observed at the relay, which makes the relay hard to decode the signal correctly and results in higher outage probability.

It can also be observed that, no matter what value of $\rho$ is, the optimal outage probability is achieved when $\theta = 0.5$. This is due to the fact that, for a symmetric system (where $P_1 = P_2$ and $d_1 = d_2$), the optimal transmit power redistribution strategy is to distribute the relay power equally.

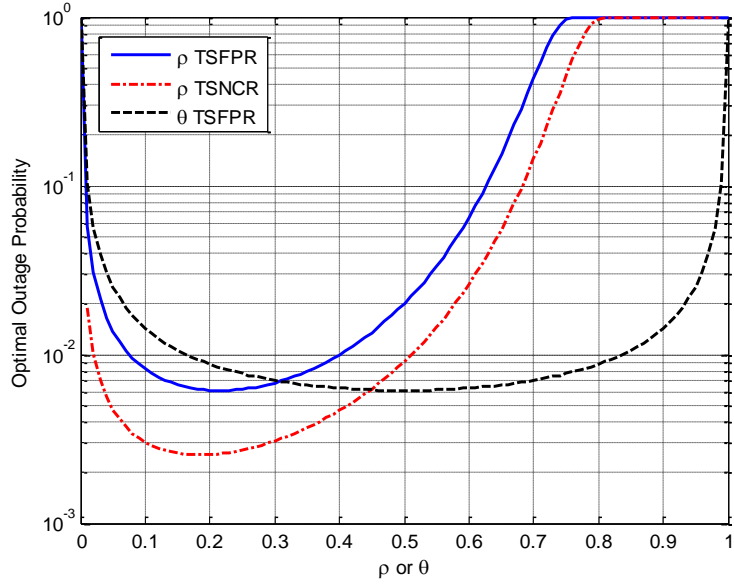

**Fig. 4** Optimal system outage probability vs $\rho$ or $\theta$

### 5.3 Effect of Relay Location on System Performance

To investigate the influence of the relay location on the optimal system outage probability and the parameters $\rho$ and $\theta$, the distance between $U_1$ and $U_2$ is set to be 2, so the distance from R to $U_2$ can be expressed as $d_2 = 2 - d_1$. **Fig. 5(a)** plots the optimal system outage probability for the two proposed protocols. It can be observed that for the TSFPR protocol, the optimal outage probability increases as $d_1$ increases, and achieve its maximum when $d_1 = d_2 = 1$. Later, the optimal outage probability starts decreasing as $d_1$ increases. As for the TSNCR protocol, the optimal outage probability first increases as $d_1$ increases, and remains unchanged when $d_1$ varies from 0.9 to 1.1, then it decreases as $d_1$ increases from 1.2. It is noteworthy that the system outage probabilities for the two proposed protocols are different from the traditional method where energy harvesting is not considered at the relay and the minimal outage probability is achieved when the relay is deployed in the middle of the two sources.

**Figure 5(b)** plots the optimal system parameters $\rho$ and $\theta$ versus $d_1$. It can be observed that for the TSFPR protocol, $\theta$ increases as $d_1$ increases, which can be easily understand: with the increase of $U_1$ to R distance, the relay needs to allocate more harvested energy to $U_1$ in order to combat the growing path loss. As for $\rho$ in TSFPR and TSNCR protocol, they both first increase and then decrease as $d_1$ increases. We also note that when $d_1 = d_2 = 1$, $\theta = 0.5$, $\rho = 0.22$ for TSFPR protocol, and $\rho = 0.19$ for the TSNCR protocol, which correspond to our previous analysis of **Fig. 4**.

**Figure 5(c)** plots the system outage capacity versus $d_1$. It can be observed that, for the two proposed protocols, both relatively lower outage capacities are achieved when the relay is deployed in the middle of the two sources. We can also find that the TSNCR protocol can greatly improve the system outage capacity compared with TSFPR.

## 5.4 Effect of Sources Transmit Power

**Figure 6** depicts the effects of two sources' transmit power on the system outage probability. We set $P_1 = 1$, and let $P_2$ vary from 0.5 to 1.5. It can be seen that, as $P_2 / P_1$ increases, both the optimal outage probabilities decrease, but the TSFPR decreases rapidly, whereas the TSNCR decreases slowly, and there is almost no change when $P_2 / P_1$ is greater than 1.4. The figure of the effects of $P_2 / P_1$ on system parameters is omitted here because the value of $P_2 / P_1$ doesn't affect $\rho$ and $\theta$ very much. No matter what value of $P_2 / P_1$ is, the optimal $\rho$ for the TSNCR is always 0.19 ($\rho$ for the TSFPR is around 0.22), and the optimal $\theta$ for the TSFPR is always 0.5, which mainly depends on the relay's position.

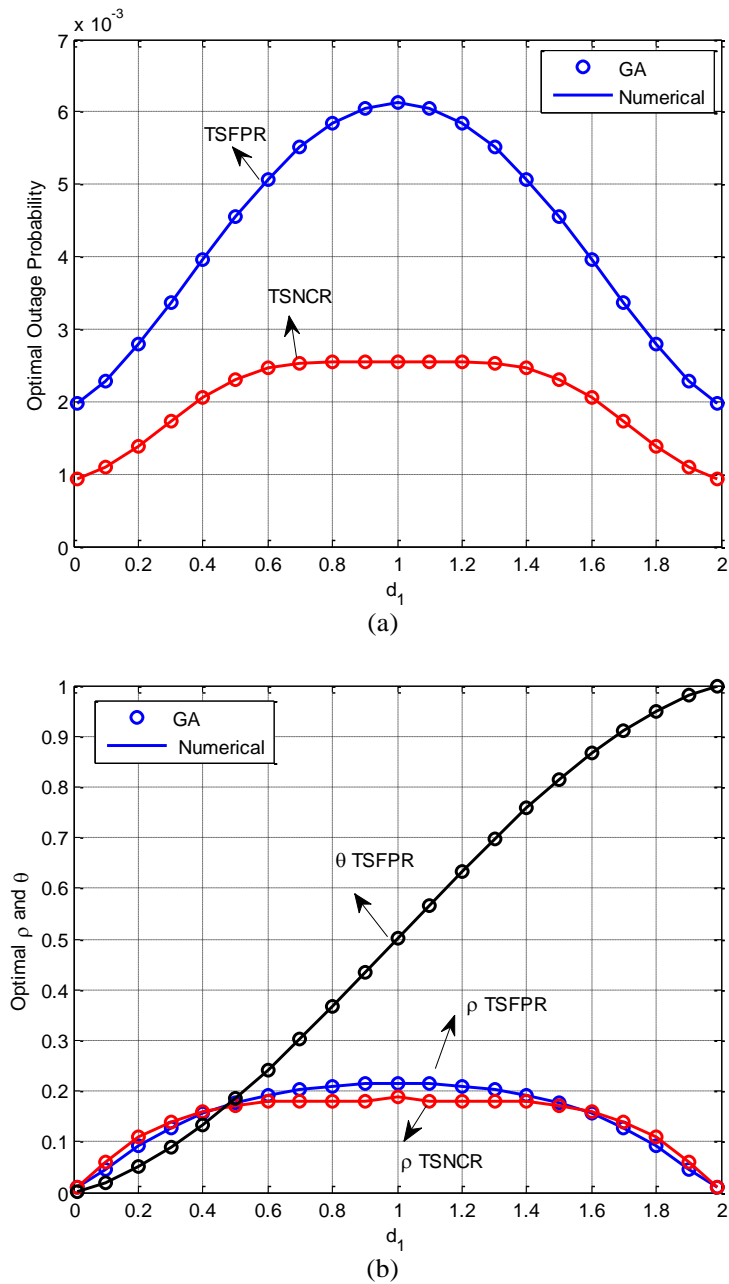

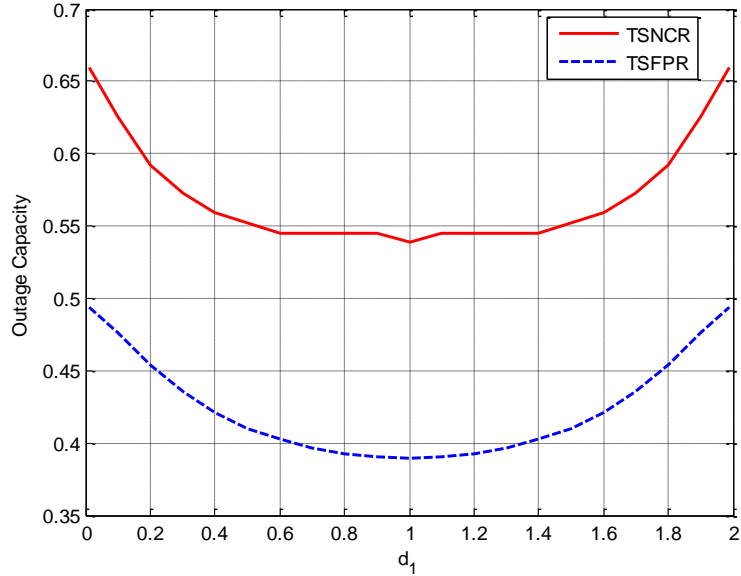

(c)

**Fig. 5** (a) Optimal system outage probability vs $d_1$ (b) optimal $\rho$ and $\theta$ vs $d_1$ (c) system outage capacity vs $d_1$.

Note that, **Fig. 6** also compares the outage performance of the two protocols: for the same sources transmit power, the TSNCR protocol outperforms the TSFPR protocol in terms of system outage probability.

The numerical results in **Fig. 5** and **Fig. 6** are obtained by computer search, which present the global optimal solution. It can be observed that the GA results perfectly match with the numerical results in **Fig. 5** and **Fig. 6**, so it implies that the proposed GA-based algorithm is able to find the global optimal solution.

### 5.5 Convergence Behavior of the Proposed GA-Based Algorithm

For an arbitrary two-way relay system with given parameters, the proposed GA-based optimization algorithm can be used to obtain the optimal $\rho$ and $\theta$ to achieve the optimal outage performance. **Fig. 7** illustrates the convergence behavior of the proposed GA-based algorithm for the TSFPR protocol. It can be seen that the algorithm converges fast, and the predefined precision is achieved within 20 runs.

## 6. Conclusion

In this paper, we studied EH and NC together in a two-way relay system and presented a TSNCR protocol. To evaluate the system performance, we derived an explicit expression of the system outage probability. In order to explore the system performance limit, we formulated an optimization problem to minimize the system outage probability. We also designed a GA-based optimization algorithm to solve the problem. Numerical results showed that in the EH two-way relay system, if NC is applied, the system outage probability can be greatly decreased. It also showed that the relay position greatly affects the system performance, where relatively worse outage performance is achieved when the relay is placed in the middle of the two sources.

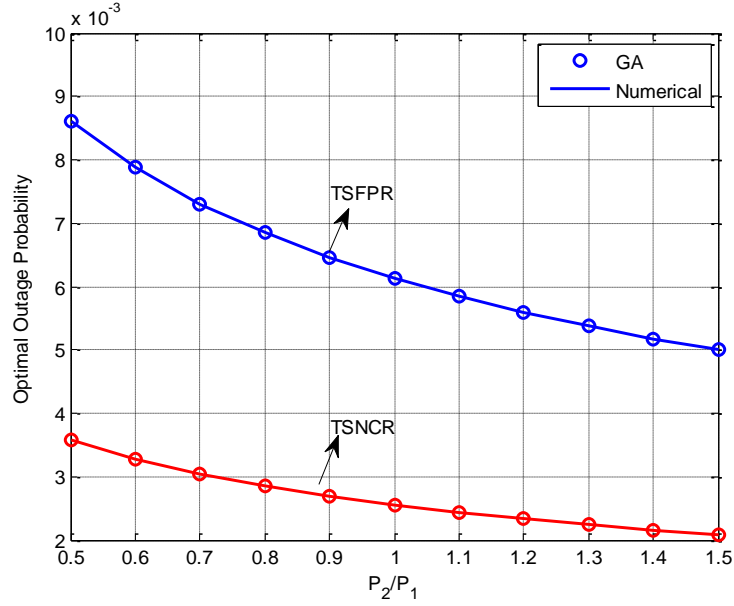

**Fig. 6** Optimal system outage probability vs $P_2/P_1$

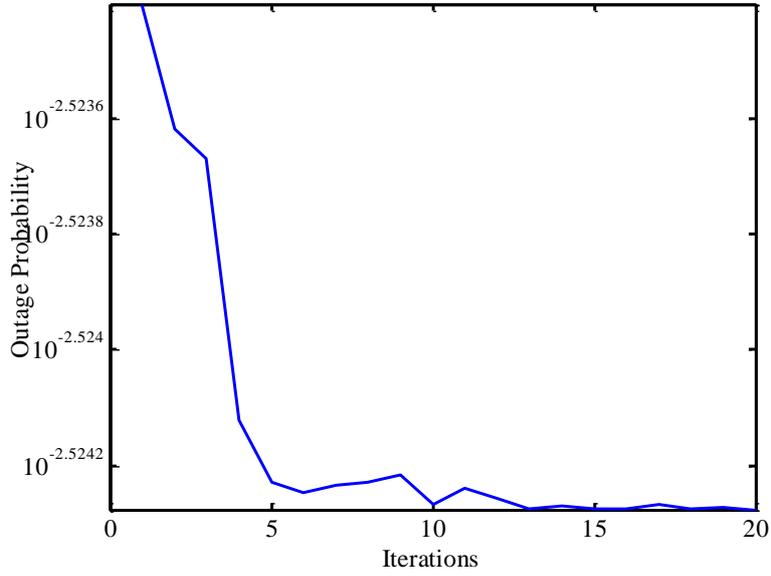

**Fig. 7** Convergence behavior of the GA-based algorithm

# Appendix A

This appendix derives the $P_{out}^{(TSNCR)}$ in (9) of *Theorem 1*.
*Proof:* Substituting (3), (4), and (7) into (8), we have that

$$P_{out}^{(TSNCR)} = 1 - \Pr[|h_1|^2 \geq a_0]\Pr[|h_2|^2 \geq b_0]\Pr[w \geq \frac{u_0}{a|h_1|^2 + b|h_2|^2}], \quad (A.1)$$

where, $w = \min(\frac{|f_1|^2}{d_1^m \sigma_{u,1}^2}, \frac{|f_2|^2}{d_2^m \sigma_{u,2}^2})$, $a = 3\rho\eta P_1 d_1^{-m}(1-\rho)^{-1}/2$, $b = 3\rho\eta P_2 d_2^{-m}(1-\rho)^{-1}/2$, $a_0 = u_0 d_1^m \sigma_{r,1}^2 P_1^{-1}$, $b_0 = u_0 d_2^m \sigma_{r,2}^2 P_2^{-1}$, $u_0 = 2^{3R_0/(1-\rho)} - 1$.

Since $|h_1|^2$ and $|h_2|^2$ are exponential random variables with mean $\lambda_{h_1}$ and $\lambda_{h_2}$, respectively, we thus have

$$\Pr[|h_1|^2 \geq a_0] = e^{-\lambda_{h_1} a_0}, \tag{A.2}$$

$$\Pr[|h_2|^2 \geq b_0] = e^{-\lambda_{h_2} b_0}. \tag{A.3}$$

By defining $z = a|h_1|^2 + b|h_2|^2$, we can write the probability density function (PDF) of $z$ as

$$f_z(z) = \begin{cases} \frac{\lambda_{h_1} \lambda_{h_2}}{a\lambda_{h_2} - b\lambda_{h_1}} (e^{-\frac{\lambda_{h_1}}{a}z} - e^{-\frac{\lambda_{h_2}}{b}z}), & a\lambda_{h_2} \neq b\lambda_{h_1} \\ \frac{\lambda_{h_1} \lambda_{h_2}}{ab} e^{-\frac{\lambda_{h_2}}{b}z} \cdot z, & a\lambda_{h_2} = b\lambda_{h_1} \end{cases} \tag{A.4}$$

Consequently, the cumulative distribution function (CDF) of $w$ is given by

$$F_w(w) = 1 - e^{-(d_1^m \sigma_{u,1}^2 \lambda_{f_1} + d_2^m \sigma_{u,2}^2 \lambda_{f_2})w}, \tag{A.5}$$

thus,

$$\Pr[w \geq \frac{u_0}{a|h_1|^2 + b|h_2|^2}] = \int_0^\infty e^{-\frac{(d_1^m \sigma_{u,1}^2 \lambda_{f_1} + d_2^m \sigma_{u,2}^2 \lambda_{f_2})u_0}{z}} \cdot f_z(z) dz$$

$$= \begin{cases} \frac{2\lambda_{h_1} \lambda_{h_2}}{a\lambda_{h_2} - b\lambda_{h_1}} (\sqrt{\frac{e_0 u_0 a}{\lambda_{h_1}}} K_1(2\sqrt{\frac{e_0 u_0 \lambda_{h_1}}{a}}) - \sqrt{\frac{e_0 u_0 b}{\lambda_{h_2}}} K_1(2\sqrt{\frac{e_0 u_0 \lambda_{h_2}}{b}})), & a\lambda_{h_2} \neq b\lambda_{h_1} \\ \frac{2\lambda_{h_1} e_0 u_0}{a} K_2(2\sqrt{\frac{e_0 u_0 \lambda_{h_2}}{b}}), & a\lambda_{h_2} = b\lambda_{h_1} \end{cases} \tag{A.6}$$

where $e_0 = e^{-\lambda_{h_1} a_0 - \lambda_{h_2} b_0}$, $K_1(\cdot)$ and $K_2(\cdot)$ denote the *first-order and second-order modified Bessel function of the second kind* respectively, and the last equality is obtained by applying the formula, $\int_0^\infty \exp(-\frac{\beta}{4x} - \gamma x) dx = \sqrt{\frac{\beta}{\gamma}} K_1(\sqrt{\beta\gamma})$ and $\int_0^\infty x \cdot \exp(-\beta x - \frac{\gamma}{x}) dx = \frac{2\gamma}{\beta} K_2(2\sqrt{\beta\gamma})$ [26].

Substituting (A.2), (A.3) and (A.6) into (A.1), we can obtain (9). This ends the proof for *Theorem 1*. ∎

## Appendix B

This appendix derives the $P_{out}^{(TSFPR)}$ in (14) of *Theorem 2*.

*Proof:* As all the channels are independent, $P_{out}^{(TSFPR)}$ in (13) can be rewritten as:

$$P_{out} = 1 - \Pr[R_{r,1} \geq R_0] \cdot \Pr[R_{u,2} \geq R_0] \cdot \Pr[R_{r,2} \geq R_0] \cdot \Pr[R_{u,1} \geq R_0]. \tag{B.1}$$

Substituting (10), (11), and (12) into (B.1), we have that

$$P_{out} = 1 - \Pr[|h_1|^2 \geq a_0] \Pr[|f_2|^2 \geq \frac{c_0}{a|h_1|^2 + b|h_2|^2}] \Pr[|h_2|^2 \geq b_0] \Pr[|f_1|^2 \geq \frac{d_0}{c|h_1|^2 + d|h_2|^2}], \quad (B.2)$$

where, $a = 2(1-\theta)\rho\eta P_1(1-\rho)^{-1}d_1^{-m}$, $b = 2(1-\theta)\rho\eta P_2(1-\rho)^{-1}d_2^{-m}$, $c = 2\theta\rho\eta P_1(1-\rho)^{-1}d_1^{-m}$, $d = 2\theta\rho\eta P_2(1-\rho)^{-1}d_2^{-m}$, $a_0 = (2^{4R_0/(1-\rho)} - 1)d_1^m \sigma_{r,1}^2 P_1^{-1}$, $b_0 = (2^{4R_0/(1-\rho)} - 1)d_2^m \sigma_{r,2}^2 P_2^{-1}$, $u_0 = 2^{4R_0/(1-\rho)} - 1$, $c_0 = u_0 d_2^m \sigma_{u,2}^2$, $d_0 = u_0 d_1^m \sigma_{u,1}^2$.

Taking steps similar to Appendix A, we can obtain (B.3), (B.4), (B.5) and (B.6),

$$\Pr[|h_1|^2 \geq a_0] = e^{-\lambda_{h_1} a_0}, \quad (B.3)$$

$$\Pr[|h_2|^2 \geq b_0] = e^{-\lambda_{h_2} b_0}. \quad (B.4)$$

$$\Pr[|f_2|^2 \geq \frac{c_0}{a|h_1|^2 + b|h_2|^2}] = \int_0^\infty e^{-\frac{\lambda_{f_2} c_0}{z}} \cdot f_z(z)dz$$

$$= \begin{cases} \frac{2\lambda_{h_1}\lambda_{h_2}}{a\lambda_{h_2} - b\lambda_{h_1}}(\sqrt{\frac{c_0 a \lambda_{f_2}}{\lambda_{h_1}}} K_1(2\sqrt{\frac{c_0 \lambda_{f_2} \lambda_{h_1}}{a}}) - \sqrt{\frac{c_0 b \lambda_{f_2}}{\lambda_{h_2}}} K_1(2\sqrt{\frac{c_0 \lambda_{f_2} \lambda_{h_2}}{b}})), & a\lambda_{h_2} \neq b\lambda_{h_1} \\ \frac{2\lambda_{h_1}\lambda_{f_2} c_0}{a} K_2(2\sqrt{\frac{c_0 \lambda_{h_2}\lambda_{f_2}}{b}}), & a\lambda_{h_2} = b\lambda_{h_1} \end{cases} \quad (B.5)$$

$$\Pr[|f_1|^2 \geq \frac{d_0}{c|h_1|^2 + d|h_2|^2}] = \begin{cases} \frac{2\lambda_{h_1}\lambda_{h_2}}{c\lambda_{h_2} - d\lambda_{h_1}}(\sqrt{\frac{d_0 c \lambda_{f_1}}{\lambda_{h_1}}} K_1(2\sqrt{\frac{d_0 \lambda_{h_1}\lambda_{f_1}}{c}}) - \sqrt{\frac{d_0 d \lambda_{f_1}}{\lambda_{h_2}}} K_1(2\sqrt{\frac{d_0 \lambda_{h_2}\lambda_{f_1}}{d}})), & c\lambda_{h_2} \neq d\lambda_{h_1} \\ \frac{2\lambda_{h_1}\lambda_{f_2} d_0}{c} K_2(2\sqrt{\frac{d_0 \lambda_{f_1}\lambda_{h_2}}{d}}), & c\lambda_{h_2} = d\lambda_{h_1} \end{cases} \quad (B.6)$$

where $K_1(\cdot)$ and $K_2(\cdot)$ are explained below (A.6). Substituting (B.3), (B.4), (B.5) and (B.6) into (B.2), we can obtain (14). This ends the proof for Theorem 2. ∎